\documentclass{iopart}

\usepackage{lineno,hyperref}
\usepackage{graphicx}
\usepackage{tikz}
\usetikzlibrary{decorations.pathmorphing}
\usepackage[T1]{fontenc}
\usepackage{booktabs} % Enhanced tables
\usepackage{mathrsfs}
\usepackage{miller}
\usepackage{amssymb}
\usepackage{amsfonts}
\usepackage{xspace}
\usepackage{braket}
\usepackage{siunitx}
\usepackage{lineno}
\usepackage{hyperref} 
\usepackage{multirow}
\usepackage{bigdelim}
\usepackage{tabularx}
\modulolinenumbers[5]

\DeclareSIUnit\electrons{e\textsuperscript{-}}
\DeclareSIUnit\neutrons{neutrons}
\DeclareSIUnit\ppm{ppm}
\DeclareSIUnit\ppb{ppb}
\DeclareSIUnit\lines{l}
\DeclareSIUnit{\calorie}{cal}

\newcommand{\Nfif}{\ensuremath{^{15}\mathrm{N}}\xspace}
\newcommand{\Nfour}{\ensuremath{^{14}\mathrm{N}}\xspace}
\newcommand{\NfifPtwoNeutral}{\ensuremath{^{15}\mathrm{N}_{3}\mathrm{V}^{0}}\xspace}

\newcommand{\NfourPtwoNeutral}{\ensuremath{^{14}\mathrm{N}_{3}\mathrm{V}^{0}}\xspace}

\newcommand{\NfifNVNminus}{\ensuremath{\mathrm{^{15}N_{2}V^{-}}}}

\newcommand{\PtwoCons}{\ensuremath{\mathrm{N}_{3}\mathrm{V}}\xspace}
\newcommand{\PtwoConsNeutral}{\ensuremath{\mathrm{N}_{3}\mathrm{V}^{0}}\xspace}
\newcommand{\PtwoConsMinus}{\ensuremath{\mathrm{N}_{3}\mathrm{V}^{-}}\xspace}

\newcommand{\NVNnb}{\ensuremath{\mathrm{N_{2}V}}\xspace} 
\newcommand{\NVnb}{\ensuremath{\mathrm{NV}}\xspace}

\newcommand{\NVminus}{\ensuremath{\NVnb^{-}}\xspace}
\newcommand{\NVneutral}{\ensuremath{\NVnb^{0}}\xspace}

\newcommand{\NVNminus}{\ensuremath{\mathrm{N_{2}V}^{-}}\xspace}
\newcommand{\NVNneutral}{\ensuremath{\mathrm{N_{2}V}^{0}}\xspace}

\newcommand{\Ns}{\ensuremath{\mathrm{N_{s}}}\xspace}

\newcommand{\Nsneutral}{\ensuremath{\mathrm{N_{s}}^{0}}\xspace}
\newcommand{\NsPlus}{\ensuremath{\mathrm{N_{s}}^{+}}\xspace}
\newcommand{\NfourNSub}{\ensuremath{^{14}\mathrm{N_{s}}^{0}}\xspace}

\newcommand{\NfifNSub}{\ensuremath{^{15}\mathrm{N}_{\mathrm{s}}^{0}}\xspace}
\newcommand{\NfifNSubPlus}{\ensuremath{^{15}\mathrm{N}_{\mathrm{s}}^{+}}\xspace}

\newcommand{\Trigonal}{\ensuremath{\mathrm{C}_{\mathrm{3v}}}\xspace}
\newcommand\scdot{{\mkern 2mu\cdot\mkern 2mu}}
\newcommand\vecto[1]{\ensuremath{\mathbf{#1}}}
\begin{document}

%\begin{frontmatter}

\title{Electron paramagnetic resonance and photochromism  of \PtwoConsNeutral{} in diamond}

\author{B.\ L.\ Green, B.\ G.\ Breeze, M.\ E.\ Newton}
\address{Department of Physics, University of Warwick, Coventry, CV4 7AL,  United Kingdom}
%\address{$^{2}$De Beers Technologies, Maidenhead, Berkshire, SL6 6JW, United Kingdom}

%\author{B.\ L.\ Green$^{1}$, B.\ G.\ Breeze$^{1}$, M.\ E.\ Newton$^{1}$, B.\ L.\ Cann$^{2}$ and D. Fisher$^{2}$}
%\address{$^{1}$Department of Physics, University of Warwick, Coventry, CV4 7AL,  United Kingdom}
%\address{$^{2}$De Beers Technologies, Maidenhead, Berkshire, SL6 6JW, United Kingdom}
\ead{m.e.newton@warwick.ac.uk}

\begin{abstract}
The defect in diamond formed by a vacancy surrounded by three nearest-neighbor nitrogen atoms and one carbon atom, \PtwoCons{}, is found in $\approx98\%$ of natural diamonds. Despite $\mathrm{N}_{3}\mathrm{V}^{0}$ being the earliest electron paramagnetic resonance spectrum observed in diamond, to date no satisfactory simulation of the spectrum for an arbitrary magnetic field direction has been produced due to its complexity. In this work, \PtwoConsNeutral is identified in \Nfif-doped synthetic diamond following irradiation and annealing. The \NfifPtwoNeutral{} spin Hamiltonian parameters are revised and used to refine the parameters for \NfourPtwoNeutral, enabling the latter to be accurately simulated and fitted for an arbitrary magnetic field direction. Study of \NfifPtwoNeutral{} under excitation with green light indicates charge transfer between \PtwoCons{} and \Ns{}. It is argued that this charge transfer is facilitated by direct ionization of \PtwoConsMinus{}, an as-yet unobserved charge state of \PtwoCons{}.
\end{abstract}
%\vspace{2pc}
%\noindent{\it Keywords}: Synthetic diamond, Defect characterization, N3, N3V\\
%\submitto{\JPCM}
\maketitle
%\end{frontmatter}
\ioptwocol
%\linenumbers
\section{Introduction}
Through careful advances in both processing and particularly synthesis, diamond has become a material with a great variety of technological applications including magnetic bio-imaging \cite{LeSage2013,Barry2016}, ultra-hard tooling \cite{Sumiya2012} and particle detectors \cite{Zamboni2013}. Underpinning these technological applications have been advances in understanding of defects and impurities within diamond, and their behaviour during synthesis and processing. The most studied point defect in diamond is the negatively-charged nitrogen-vacancy centre \NVminus{} (due to its superlative optical and spin properties \cite{Doherty2013,Rondin2014}), a defect which belongs to the $\mathrm{N}_n\mathrm{V}$ family of defects, with $n=\numrange{1}{4}$. Electron paramagnetic resonance (EPR) and optical signatures have been identified for $\mathrm{NV^{0/-}}$ \cite{Felton2008,Felton2009a}, $\mathrm{N_{2}V^{0/-}}$ \cite{Loubser1981,Green2015}, $\mathrm{N_{3}V^{0}}$ \cite{Davies1978,VanWyk1993}, and optical-only for $\mathrm{N_{4}V^{0}}$ (known as the B-center) \cite{Collins2003a}. %Indirect evidence exists for \PtwoConsMinus{} \cite{Mita1997b} and $\mathrm{NV^{+}}$ \cite{Pfender2016}.

\begin{figure}[ht]
	\centering
	\includegraphics[width=\columnwidth]{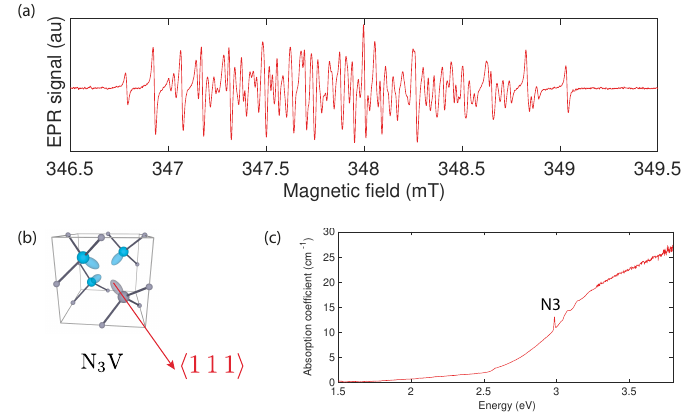}
	\caption{(a) EPR spectrum of a diamond containing \NfourPtwoNeutral{}, with the applied magnetic field $B\|\hkl<001>$. The complex structure is due to electron-nuclear interactions with the three $I=1$ \Nfour{} nuclei. (b) Atomic structure of \PtwoCons{} in diamond, with the \hkl<111> \Trigonal{} symmetry axis highlighted. (c) UV-Vis absorption spectrum of a synthetic \Nfif{}-doped HPHT-grown diamond containing \PtwoConsNeutral{} (ZPL at \SI{2.988}{\electronvolt}) and substitutional nitrogen (\Nsneutral{}, absorption ramp starting at approximately \SI{2}{\electronvolt}). }
	\label{fig:structure_and_spectra}
\end{figure}

The N3 absorption and luminescence band, with a zero-phonon line (ZPL) at \SI{2.988}{\electronvolt} (\SI{415}{\nano\meter}, Fig.~\ref{fig:structure_and_spectra}(c)), is commonly observed in natural and treated synthetic diamonds. Indeed, its ubiquity in natural diamond has led to the development of a commercial categorization device which looks for the presence of the absorption band \cite{Welbourn1996}. A complex EPR spectrum, (known as P2, Fig.~\ref{fig:structure_and_spectra}(a)), is associated with the N3 absorption band by correlation between the EPR and optical signatures \cite{Davies1978}. Through careful electron nuclear double resonance (ENDOR) experiments, the EPR spectrum was determined to arise at a center containing three nitrogen atoms decorating a vacancy, \PtwoConsNeutral{} (Fig.~\ref{fig:structure_and_spectra}(b)) \cite{VanWyk1982,VanWyk1993}. This structure has \Trigonal{} symmetry and hence there are four equivalent orientations of the defect within the diamond lattice, each defined by its unique \hkl<111> axis. Each nitrogen atom is covalently bonded to its three carbon nearest-neighbors, with its lone pair directed approximately into the vacancy. We therefore expect the unpaired electron probability density to reside primarily on the carbon atom nearest-neighbor to the vacancy, with a commensurately small hyperfine interaction with each of the nitrogen constituents. 

\begin{table*}[btp]
	\centering
	\caption[The measured spin Hamiltonian parameters for the \PtwoConsNeutral defect in \Nfif-doped diamond]{The measured spin Hamiltonian parameters for the \PtwoCons defect. Parameters have been given for \Nfour-doped diamond by scaling the measured \Nfif hyperfine values by the ratio of the nuclear g-values $g_{14}:g_{15}$ ($g_{14} = \num{0.403761}$ and $g_{15} = \num{-0.566378}$ \cite{Stone2005}). The g-tensor values are calculated assuming that \NfourNSub has an isotropic g-value of \num{2.0024}. $\theta$ measured from \hkl[110] toward \hkl[001]; quadrupolar rhombicity given as $\eta = (P_{x} - P_{y})/P_{z}$. Blanks in \NfifPtwoNeutral{} parameters indicate values are as \NfourPtwoNeutral.}
	\label{tab:spin_hamiltonian_parameters}
	\setlength{\tabcolsep}{0.1em}
	\centering
	\small
	\begin{tabularx}{\textwidth}{lXllS[tight-spacing=true]XS[tight-spacing=true]S[tight-spacing=true]S[tight-spacing=true]S[tight-spacing=true]XS[tight-spacing=true]S[tight-spacing=true]S[tight-spacing=true]}
		\toprule
			& 	& \multicolumn{3}{c}{Zeeman} & & \multicolumn{4}{c}{Hyperfine} & & \multicolumn{3}{c}{Quadrupole}\\
			& 	& 	\multicolumn{1}{c}{$g_{\|}$} & 	\multicolumn{1}{c}{$g_{\perp}$} & \multicolumn{1}{c}{$\theta$(\si{\degree})} & & \multicolumn{1}{c}{$A_{1}$}& \multicolumn{1}{c}{$A_{2}$}& \multicolumn{1}{c}{$A_{3}$} & \multicolumn{1}{c}{$\theta$(\si{\degree})} & & \multicolumn{1}{c}{$P_{\|}$}& \multicolumn{1}{c}{$\eta$} & \multicolumn{1}{c}{$\theta$(\si{\degree})}\\
		\midrule
		\multicolumn{1}{l}{\NfourPtwoNeutral{} \cite{VanWyk1993}}&&2.0023(2)&2.0032(2)&35.26&&7.4+-0.1&7.4+-0.1&11.2+-0.1&158+-1&&-4.8+-0.1&0&145+-2\\
		\multicolumn{1}{l}{\NfourPtwoNeutral{} (this work)}&&2.00241(5)&2.00326(5)&35.26&&7.44+-0.04&7.46+-0.04&11.30+-0.04&157.8+-0.2&&-4.73+-0.05&0&144.7+-0.5\\
		\multicolumn{1}{l}{\NfifPtwoNeutral{} (this work)}&&&&&&10.44+-0.05&10.46+-0.05&15.85+-0.05&&&\multicolumn{3}{c}{--- N/A ---}\\
		\bottomrule
	\end{tabularx}
\end{table*}

The spin Hamiltonian for a single electron, multiple nucleus system is given by
\begin{equation*}
	\vecto{H} = \mu_B \vecto{B}^{T} \scdot \vecto{g} \scdot \vecto{S} + \sum_{i}^{N} \vecto{S}^{T} \scdot \vecto{A}_{i} \scdot \vecto{I}_{i} + \vecto{I}^{T}_{i} \scdot \vecto{Q} \scdot \vecto{I}_{i} \;,
\end{equation*}
with $i$ summed over the nuclei. These terms represent the electronic Zeeman, electron-nuclear hyperfine, and nuclear quadupole interactions, respectively. Applying a magnetic field lifts the degeneracy of the electronic states $m_S$ via the Zeeman interaction, with transitions between the states driven by resonant high frequency magnetic fields (usually in the microwave region) --- the electron paramagnetic resonance phenomenon. Each nucleus splits the electron resonance line into $2I + 1$ lines, where $I$ is the nuclear spin of the isotope in question --- $1$ for \Nfour{}; $1/2$ for \Nfif{}. For an arbitrary magnetic field $\vecto{B}$ applied to \PtwoConsNeutral{} we therefore expect $3^{3} = 27$ ($2^{3} = 8$) lines per symmetry-related orientation for the \Nfour{} (\Nfif{}) case. The \Nfour{} case is further complicated by the quadrupolar interaction, which is only non-zero for $I>1/2$: this interaction enables so-called ``forbidden'' ($\Delta m_S = 1$; ${\Delta m_I \neq 0}$) transitions to acquire appreciable intensity, and can yield up to $27^2 = 729$ lines per symmetry-related orientation. 

Due to the complexity of the \NfourPtwoNeutral{} EPR spectrum, very small changes to the spin Hamiltonian parameters have a dramatic effect on the quality of a simulation of the measured spectrum. The published spin Hamiltonian parameters (see Table~\ref{tab:spin_hamiltonian_parameters}) produce a satisfactory simulation only for an external magnetic field $B\|\hkl<001>$. For all other directions, the published parameters produce a poor simulation (Fig.\ref{fig:n14_spectra}, bottom): this presents a problem when attempting to quantify defect concentrations in crystals whose orientation is not known, or where experimental geometric constraints prevent the crystal being oriented along \hkl<100>. It is clear that the production of \PtwoConsNeutral{} in a \Nfif{}-doped diamond would greatly simplify the observed EPR spectrum and enable revised spin Hamiltonian parameters to be obtained. 

Exploiting advances in synthesis, we have used \Nfif{}-doped synthetic diamond to determine more accurate spin Hamiltonian parameters for the \PtwoConsNeutral{} EPR center. Additionally, we have studied the photochromism behaviour of \NfifPtwoNeutral{} under illumination with green light.%, and used them to obtain a higher precision value for dipole oscillator strength for the \SI{415}{\nano\meter} absorption transition at the same center. 

\section{Experimental}
Atmospheric nitrogen is readily incorporated into diamond during high pressure high temperature (HPHT) growth if chemical nitrogen traps (``getters'') are not employed. The sample used in this experiment was grown using a HPHT growth capsule that was outgassed under vacuum and subsequently backfilled with an isotopically enriched $\mathrm{N_{2}}$ gas \cite{Stromann2006} with a nitrogen isotopic ratio of \Nfour{}:\Nfif{} of approximately 5:95. A total substitutional nitrogen concentration of \SI{84+-3}{\ppm} was measured after growth. The sample was irradiated to a total dose of approximately \SI{5E17}{\neutrons\per\centi\metre\squared} under atmospheric conditions at an estimated sample temperature of approximately \SI{200}{\celsius} \cite{Liggins2010b}. The sample was then annealed for \SI{15}{\hour} at \SI{1500}{\celsius} in a non-oxidizing atmosphere, before being annealed further under HPHT conditions for \SI{1}{\hour} at a nominal temperature of \SI{1900}{\celsius}.

After processing, the sample was measured (by IR absorption) to contain $\NfifNSub{} \approx \SI{20}{\ppm}$; $\NfifNSubPlus{} \approx \SI{5}{\ppm}$; and $\mathrm{N_{2}^0} \approx \SI{40}{\ppm}$ (known as A-centers). The concentrations are approximate as we do not have a suitable reference spectrum for $^{15}\mathrm{N_4V}^0$: we estimate approximately \SI{15}{\ppm} nitrogen in $\mathrm{N_4V}^0$ form by fitting a $^{14}\mathrm{N_4V}^{0}$ spectrum to the IR data. The total concentration of \NfifPtwoNeutral generated by this processing regime was approximately \SI{1.6+-0.2}{\ppm}, as measured by EPR. Photoluminescence (PL) measurements of the sample post-processing were dominated by \PtwoConsNeutral{}, \NVNneutral{}, \NVNminus{}, \NVneutral{} and \NVminus{}. The high levels of nitrogen aggregation at a moderate annealing temperature (\SI{1900}{\celsius}) are ascribed to the abundance of vacancies introduced by the neutron irradiation \cite{Collins1980}.

\section{Results}

\subsection{Spin Hamiltonian}
\begin{figure}[htb]
	\centering
	\includegraphics[width=\columnwidth]{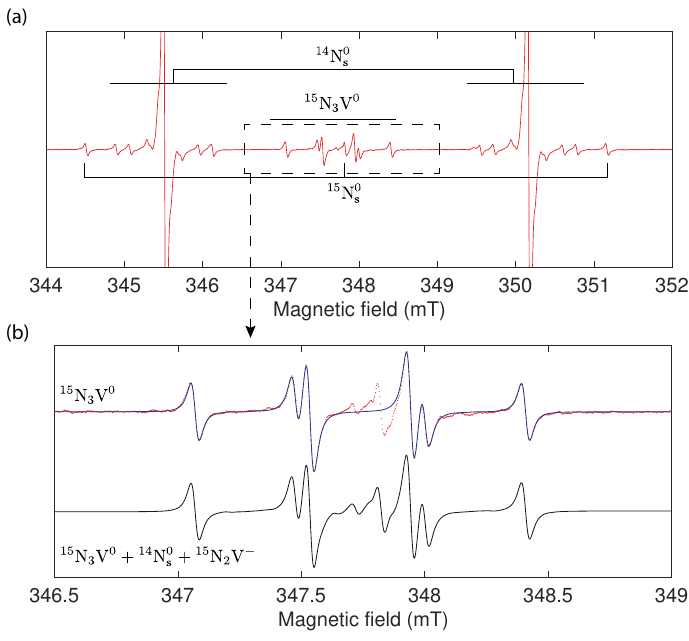}
	\caption{(a) Measured EPR spectrum of synthetic \Nfif{}-doped HPHT-treated sample with applied magnetic field $B\|\hkl<001>$. Three paramagnetic systems are visible and labeled. The \NfifPtwoNeutral EPR spectrum is significantly simpler than the \Nfour{} analogue (Fig.~\ref{fig:structure_and_spectra}(a)). (b) Zoom of marked area in (a). Top: Experimental spectrum is dotted, generated \NfifPtwoNeutral{} spectrum is the solid line. Bottom: Generated spectrum also including \NfourNSub{} and \NfifNVNminus{}, to account for central resonances in experimental spectrum.}
	\label{fig:n15_spectra}
\end{figure}

A typical EPR spectrum of the sample with an applied magnetic field $\vecto{B}\|\hkl<001>$ is given in figure~\ref{fig:n15_spectra}. \NfifPtwoNeutral{} was identified by modifying the published Hamiltonian parameters for \NfourPtwoNeutral{}: the hyperfine interaction strengths were scaled by the ratio of the isotopic nuclear g-values ($g_{14} = \num{0.403761}$ and $g_{15} = \num{-0.566378}$ \cite{Stone2005}) and the quadrupolar interaction was removed.

Unlike the \Nfour{} spectrum (Fig.~\ref{fig:structure_and_spectra}(a)), the interpretation of the \NfifPtwoNeutral{} spectrum is relatively simple (figure~\ref{fig:n15_spectra}). As a consequence of the \Trigonal{} symmetry of the defect, all four symmetry-related orientations are equivalent when the external magnetic field is applied along \hkl<001> and only one $g$-value is observed. Furthermore, the hyperfine interaction for each nitrogen nucleus is also identical, and for three equivalent $I=1/2$ nuclei at an $S=1/2$ centre, a 1:3:3:1 intensity pattern is expected. In fact, the approximate pattern is 1:1:2:2:1:1, with the highest-intensity lines split by second-order hyperfine effects. As expected, the spectrum is significantly simpler than in the \Nfour{} case.

Further spectra were measured with the applied magnetic field along \hkl<110> and \hkl<111>, in addition to \hkl<001>. The spectra for all three orientations were fitted simultaneously: the obtained spin Hamiltonian parameters are given in Table~\ref{tab:spin_hamiltonian_parameters}. The revised \NfifPtwoNeutral{} spin Hamiltonian parameters were then used to improve the fit for \NfourPtwoNeutral{}, where the only only free parameters are those relating to the quadrupole interaction. 

\begin{figure}
	\centering
	\includegraphics[width=\columnwidth]{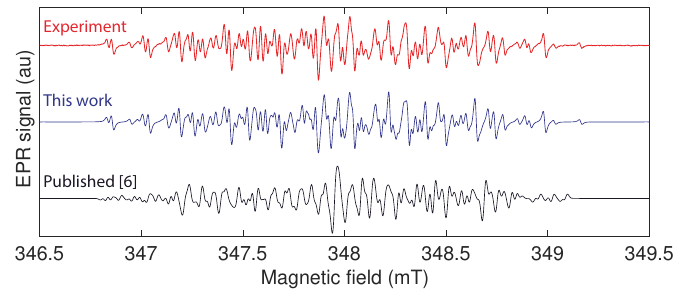}
	\caption{EPR spectrum of a natural diamond with applied magnetic field $B\|\hkl<111>$. Top to bottom: experimental data; spectrum generated using the spin Hamiltonian parameters given in table~\ref{tab:spin_hamiltonian_parameters}; spectrum generated using the spin Hamiltonian parameters given in \cite{VanWyk1993}. The calculated \Nsneutral{} spectrum was subtracted from the experimental data before fitting.}
	\label{fig:n14_spectra}
\end{figure}

A second (natural) diamond with natural nitrogen isotope abundance (\SI{100}{\percent} \Nfour{}), and containing approximately \SI{25}{\ppm} of \NfourPtwoNeutral{} was studied. Once again, EPR spectra of three high-symmetry directions were measured and simultaneously fitted. The updated spin Hamiltonian parameters enable fitting of the \NfourPtwoNeutral{} spectrum along arbitrary directions (see figure~\ref{fig:n14_spectra}). It is interesting to note that the updated parameters all lie within the quoted errors of the published data; however, the spectrum is complex due to several interactions of similar magnitudes, and hence very small relative shifts have a dramatic effect on the generated spectrum.

\subsection{Photochromism}
The negative charge states of \NVnb{} and \NVNnb{} have both been identified; however, no negative analogue of \PtwoConsNeutral{} has been identified by EPR, optical absorption or PL. Indirect evidence for \PtwoConsMinus{} was obtained in material with high \Nsneutral{} by monitoring the ZPL of \PtwoConsNeutral{} at \SI{2.988}{\electronvolt}) (\SI{415}{\nano\meter}) in absorption while pumping with a filtered arc lamp: upon illumination with light of energy \SI{>1.65}{\electronvolt} the ZPL was seen to increase \cite{Mita1997b}. This energy was interpreted as the ionization threshold for \PtwoConsMinus{}: as only the \PtwoConsNeutral ZPL was monitored, no corresponding donor (or acceptor) was identified. No effect was observed when the same experiment was performed in material with low single nitrogen concentration. 

The visible absorption of \Nsneutral{} is a broad ramp starting at approximately \SI{2.0}{\electronvolt} and increasing in intensity toward the ultraviolet (see underlying ramp of figure~\ref{fig:structure_and_spectra}(c)) with a small feature at \SI{4.59}{\electronvolt} (\SI{270}{\electronvolt}) \cite{Khan2009}. The absorbance of this sample at \SI{>4.0}{\electronvolt} saturates our spectrometer, and hence any change in \Nsneutral{} concentration will manifest only as a small baseline offset to the ramp. Unfortunately, it is experimentally difficult to exclude any optical pump to such a degree no pump light enters the spectrometer detection path (and thus also manifests as a baseline offset). The effect of optical pumping on \Nsneutral{} and \PtwoConsNeutral{} was therefore measured separately: the \Nsneutral{} change was monitored by via its IR absorption; and \PtwoConsNeutral{} using visible absorption.

\begin{figure}
	\centering
	\includegraphics[width=\columnwidth]{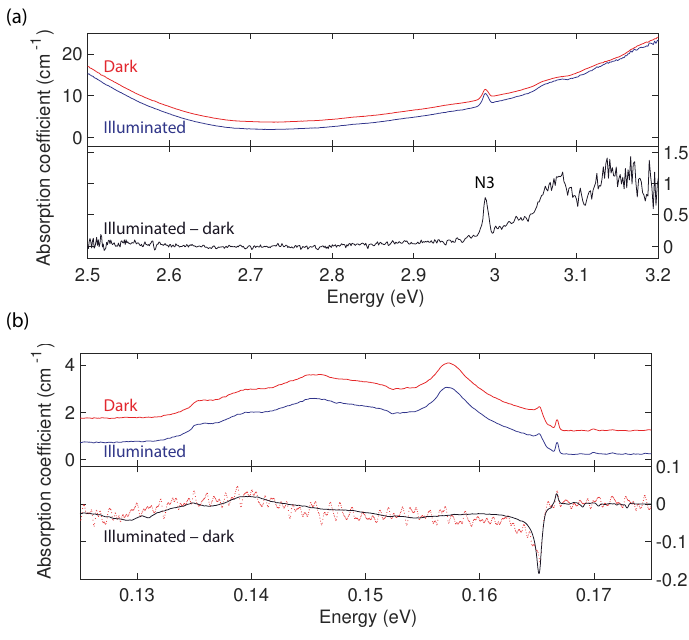}
	\caption{(a) UV-Vis absorption spectra of the \Nfif{}-doped sample at \SI{110}{\kelvin}. Top: spectra collected with and without \SI{100}{\milli\watt} illumination at \SI{2.33}{\electronvolt} (\SI{532}{\nano\meter}). Dark spectrum offset by \SI{1}{\per\centi\meter} for clarity. Bottom: difference spectrum: there is a clear increase in the N3 ZPL and associated vibronic band under illumination. (b) The one-phonon infrared absorption spectrum of the \Nfif{}-doped sample at \SI{110}{\kelvin}. Top: spectra collected with and without \SI{100}{\milli\watt} illumination at \SI{2.38}{\electronvolt} (\SI{520}{\nano\meter}). Dark spectrum offset by \SI{1}{\per\centi\meter} for clarity. Bottom: difference spectrum (experimental, dots) and fit (solid line). The fit was generated by performing least squares fitting of \Nfif{} reference spectra for \NsPlus{} and \Nsneutral{}, and represents a change of $-\SI{1.3+-0.1}{\ppm}$ and $+\SI{1.3+-0.2}{\ppm}$, respectively.}
	\label{fig:charge_transfer}
\end{figure}

Measurements of the sample at \SI{110}{\kelvin} show an increase in \PtwoConsNeutral{} of approximately \SI{0.1}{\ppm} (using the calibration in \cite{Davies1999}) when illuminated by \SI{100}{\milli\watt} at \SI{2.33}{\electronvolt} (\SI{532}{\nano\meter}). Measurements in the infrared record changes in \NsPlus{} and \Nsneutral{} of $-\SI{1.3+-0.1}{\ppm}$ and $+\SI{1.3+-0.2}{\ppm}$, respectively, when the sample is illuminated with \SI{100}{\milli\watt} at \SI{2.38}{\electronvolt} (\SI{520}{\nano\meter}). Taken together, these results suggest that illumination drives the process \begin{equation}(\PtwoConsMinus{} + \NsPlus{} \rightarrow \PtwoConsNeutral{} + \Nsneutral{})\;.\label{eqn:chTran}\end{equation} The excitation energies employed here are close to the photoionization threshold of \Nsneutral{} (approximately \SIrange{1.9}{2.2}{\electronvolt} \cite{Heremans2009,Isberg2006}). The energy for $\NsPlus{} + \nu \rightarrow \Nsneutral{} + h^{+}$ is \SI{4.0}{\electronvolt} \cite{Jones2009b} and hence we cannot drive this process in our experiments. We therefore infer that our optical excitation is leading to ionization of \PtwoConsMinus{} directly, and during illumination we are driving process~(\ref{eqn:chTran}) in both the forward and backward direction due to consecutive ionization of \PtwoConsMinus{} and \Nsneutral{}. 

The historical lack of observation of \PtwoConsMinus{} is not unexpected: \PtwoCons{} represents significant nitrogen aggregation in diamond, and it is unusual to observe high-order aggregates and single substitutional nitrogen in the same sample, as required for the initial $\Ns{} \leftrightarrow \PtwoCons{}$ process to occur. The identification of the donor in this process as \Nsneutral{} combined with previous ionization results \cite{Mita1997b} indicates that the photoionization threshold for \PtwoConsMinus{} lies in the range \SIrange{1.65}{2.2}{\electronvolt}. 

Evidently the change in \PtwoConsNeutral{} concentration cannot account for the entirity of the charge transfer. We expect the processes $(\NVNminus{} + \NsPlus{} \rightarrow \NVNneutral{} + \Nsneutral{})$) and ($\NVminus{} + \NsPlus{} \rightarrow \NVneutral{} + \Nsneutral{}$) to be occurring simultaneously: unfortunately, both charge states of NV were below absorption detection limits; and the bandblock filter used to exclude the laser light from the spectrometer also blocks the ZPL of both charge states of $\mathrm{N_2V}$. Subsequent absorption measurements in the range \SIrange{1.37}{2.18}{\electronvolt} (\SIrange{900}{570}{\nano\meter}) did not identify any sharp features with or without illumination. 
 
% \subsection{EPR - optical correlation}

% \begin{figure}
% 	\centering
% 	\includegraphics[width=\columnwidth]{graphs/epr_uv_corr/p2_n3_corr.pdf}
% 	\caption{Correlation between the integrated intensities of the P2 EPR spectrum and the N3 optical absorption transition. Systematic errors are estimated at \SI{10}{\percent}; random errors are higher (particularly in the optical measurement) due to the inhomogeneity of the samples.}
% 	\label{fig:p2_n3_corr}
% \end{figure}

\section{Conclusion}
Advances in synthesis and processing have enabled us to create a sample with a significant quantity of \NfifPtwoNeutral{}. Due to the dramatic simplification of the \Nfif spectrum we have been able to refine the spin Hamiltonian parameters of both \NfifPtwoNeutral{} and \NfourPtwoNeutral{}: this enables the fitting and hence quantification of \PtwoConsNeutral{} with the sample in an arbitrary orientation relative to the applied magnetic field. Subsequent optical absorption measurements of the \Nfif{}-doped sample under illumination have indicated charge transfer between substitutional nitrogen \Ns{} and \PtwoCons{}. The simplest model consistent with our results suggests the presence of the as-yet unobserved charge state \PtwoConsMinus{}. Careful absorption and photoluminescence measurements in the \SIrange{1.65}{2.2}{\electronvolt} range may identify features associated with \PtwoConsMinus{}. 

\section*{References}
%\bibliographystyle{model1a-num-names}
%\bibliographystyle{iopart-num}
%\bibliography{library}
\providecommand{\newblock}{}

\end{document}